\documentstyle[epsf]{l-aa}         % LaTeX A&A  Standard Fonts
\epsfverbosetrue
\newcommand{\Teff}{\mbox{$T_{\rm eff}$}}

\newcommand{\ion}[2]{\mbox{#1\ {\scriptsize #2}}}
\newcommand{\lppr}{\stackrel{<}{\scriptstyle \sim}}
\newcommand{\lappr}{\raisebox{-0.4ex}{$\lppr$}}

\def\fm{\hbox{$.\!\!^{\rm{m}}$}}

\def\fs{\hbox{$.\!\!^{\rm{s}}$}}
\def\hs{\hbox{$\hbox{\thinspace \fs}$}}

\newcount\equanumber      \equanumber=0
{\global\advance\equanumber by 1}
\def\begineq#1{\global\advance\equanumber by 1
              \begingroup $$ #1 \hfill \eqno {\rm (\number\equanumber)}}

\begin{document}
  
\thesaurus{08(08.23.1, 08.13.1, 08.09.2:HE\,1211-1707, 
08.09.2:HE\,0127-3110 , 08.09.2:HE\,2201-2250, 08.09.2:HE\,0000-3430)}

\title{Discovery of four white dwarfs with strong magnetic fields by the
Hamburg/ESO Survey
  \thanks{Based on observations at the European Southern Observatory
in Chile,  the
German-Spanish Astronomical Center, Calar Alto, Spain,
operated by the Max-Planck-Institut f\"ur Astronomie, Heidelberg, jointly with
the Spanish National Commission for Astronomy, and taken with the IUE 
observatory, Villafranca, Spain.
} }
\author{D. Reimers\inst{1}, S. Jordan\inst{2}, D. Koester\inst{2}, N. Bade
\inst{1}, Th. K\"ohler\inst{1}, L. Wisotzki\inst{1}
}

\offprints{S. Jordan}

\institute{Hamburger Sternwarte,  Gojenbergweg 112, D-21029 Hamburg, Germany
            \and
           Institut f\"ur Astronomie und Astrophysik der Universit\"at,
           D-24098 Kiel, Germany
           \newline INTERNET (SJ):  jordan@astrophysik.uni-kiel.d400.de}

\date{received Obtober 1995; accepted December 1995}

\maketitle
\markboth{D.~Reimers, S. Jordan, D. Koester et al.: Discovery of four magnetic 
white dwarfs by  the Hamburg/ESO Survey}{D. Reimers, S. Jordan, D. Koester et al.: 
Discovery of four magnetic white dwarfs  by  the Hamburg/ESO Survey}

\begin{abstract}
Four magnetic white dwarfs have been found in the course of the Hamburg/ESO
Survey for bright QSOs. The objects have  been selected as QSO candidate on
the basis of its blue continuum and the apparent absence of strong
hydrogen or helium lines.
 One star, HE\,1211-1707,  shows a rather fast spectral
%sj: second hottest deleted
variability: both the strength and the position of the shallow absorption
features  change on a time scale of 20\,minutes. We  interpret this
variability as being   due to a magnetic field on the surface of a rotating
white dwarf, having a
relatively uniform magnetic field on one hemisphere and a much larger
spread of field strengths visible during other phases of the rotational
period. All attempts to determine the magnetic field structure
in detail 
with the help of synthetic spectra have failed so far because the star
must have  a rather complicated field geometry. However, both the optical and
the UV spectra indicate that a significant  part of the surface is
dominated by  a magnetic field strength of  about 80\,MG.

The spectrum of HE\,0127-3110 is also rotationally modulated. 
This star (approximate range of magnetic fields: 85-345\,MG) as well as 
 HE\,2201-2250 (a spectroscopic  twin of HE\,0127-3110)  and
HE\,0000-3430 (43-118\,MG)
 could be reasonably well reproduced with the help of
theoretical spectra calculated assuming magnetic dipoles which are offset
by 0.1 and 0.2 stellar radii along the magnetic axis. This result is in
agreement with the assumption that  Ap stars, also showing significant
deviations from a centered dipole, are the progenitors of magnetic 
white dwarfs. 
\end{abstract}


\section{Introduction}
Degenerate stars with strong magnetic fields have been studied
intensively in recent years (see Chanmugam 1992 for a review),
 ever since Greenstein et al. (1985)
could show that in Grw$ +70^{\circ}8247$, whose peculiar absorption bands
(Minkowski, 1938) resisted for decades a successful
identification, these bands are due to hydrogen in magnetic fields
of several hundred MG. In the meantime, there has been
considerable progress in numerical quantum mechanical calculations
of the hydrogen levels in strong magnetic fields (R\"osner et al.
1984; Forster et al. 1984; Henry \& O`Connell, 1984; Wunner et al.,
1985), in applying these results to stellar atmospheres and line
spectrum synthesis calculations (Achilleos and Wickramasinghe 1989,
Putney and Jordan 1995 and references therein) and in discovering more single
degenerate
%gau\ss -> gauss
stars field strengths of several hundred Megagauss, cf. Table\,2 in
Schmidt \&\ Smith  (1995). Several of these stars like PG\,1031+234
(Schmidt et al. 1986, Latter et. al. 1987)
or PG\,1015+014 (Wickramasinghe \&\ Cropper 1988) in addition exhibit strong
rotational modulation of their spectral lines with periods as short as 1.65
hours. 
As relics of stellar cores, the study of magnetic
fields in white dwarfs should shed important light on the role
such fields play in stellar formation and main sequence evolution.
In this paper we present spectra of four newly discovered magnetic 
white dwarfs. Three of the objects have been successfully modelled with
offset dipole configurations. We also present time resolved spectra of
HE\,1211-1707, probably one of the fastest rotating white dwarfs with 
a rather complicated magnetic field configuration.

\newpage

\begin{figure}[htbp]
\fbox{{\centering
\epsfxsize=8.5cm
\epsffile{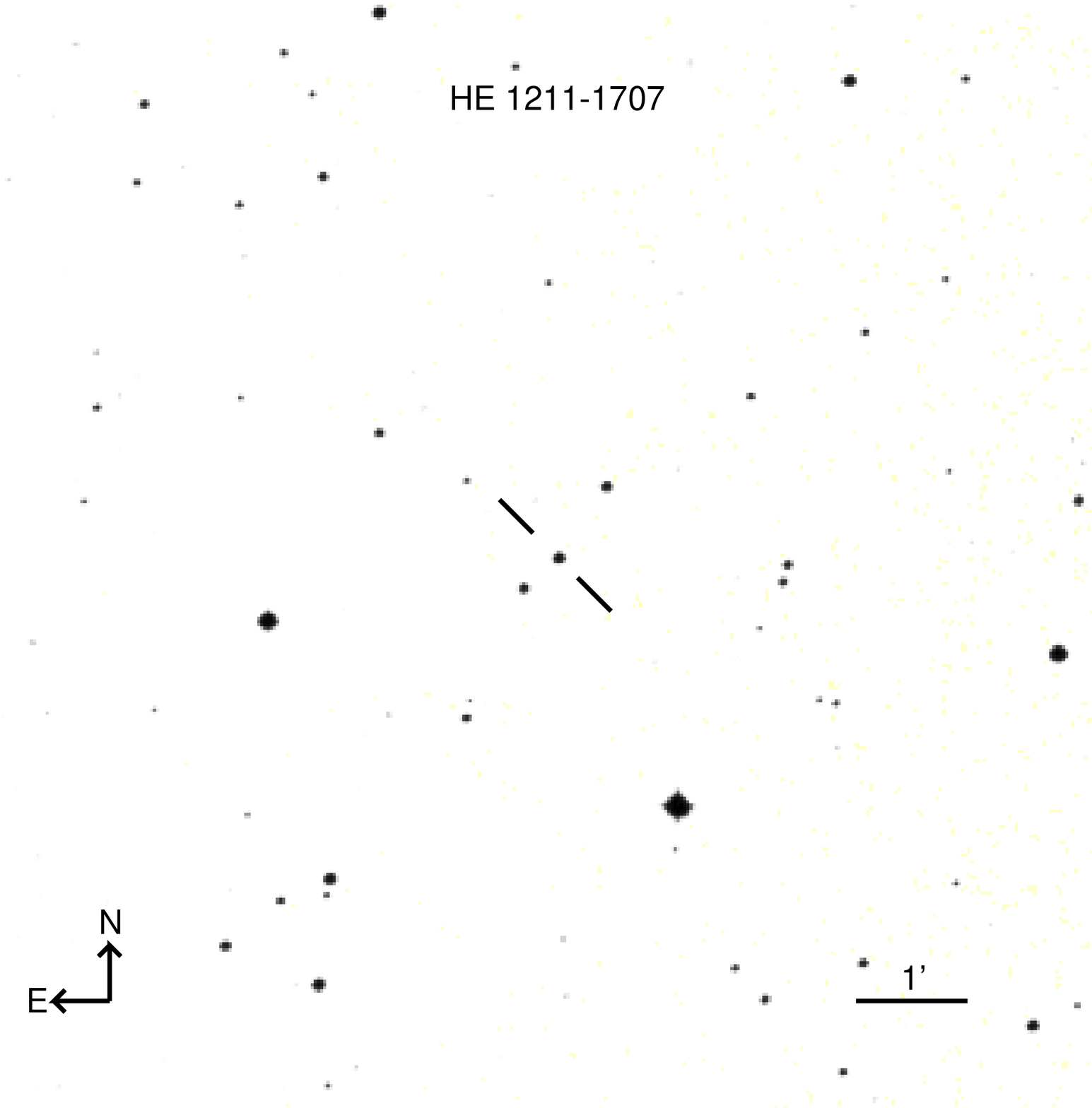}}}

\vspace{1.2mm}
\fbox{{\centering
\epsfxsize=8.5cm
\epsffile{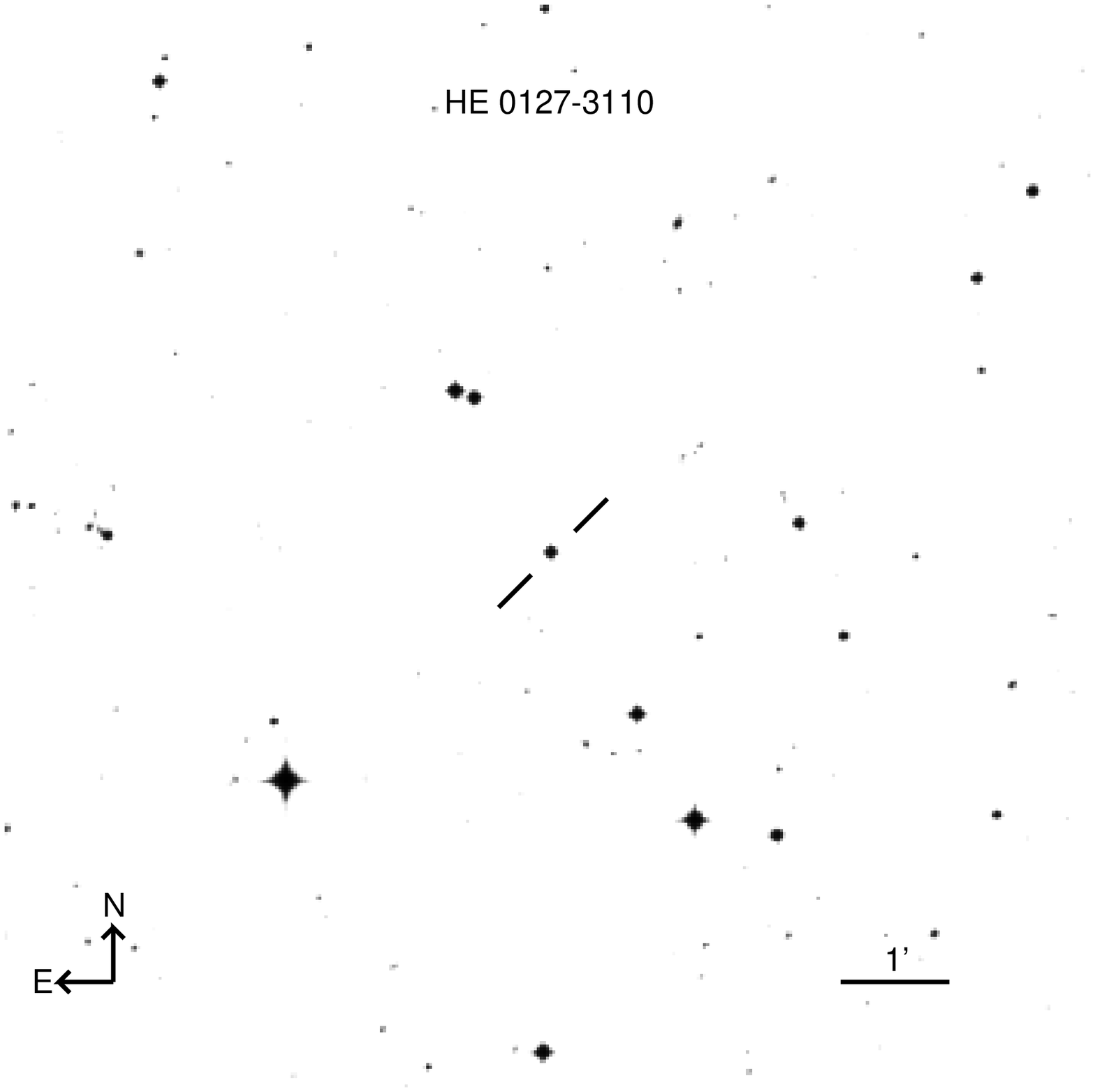}}}
\caption[]{Finding charts for HE\,1211-1707 and HE\,0127-3110 produced
with the help of the Digitized Sky Survey
}
\label{ffc1}
\end{figure}  

\begin{figure}[htbp]
\fbox{{\centering
\epsfxsize=8.5cm
\epsffile{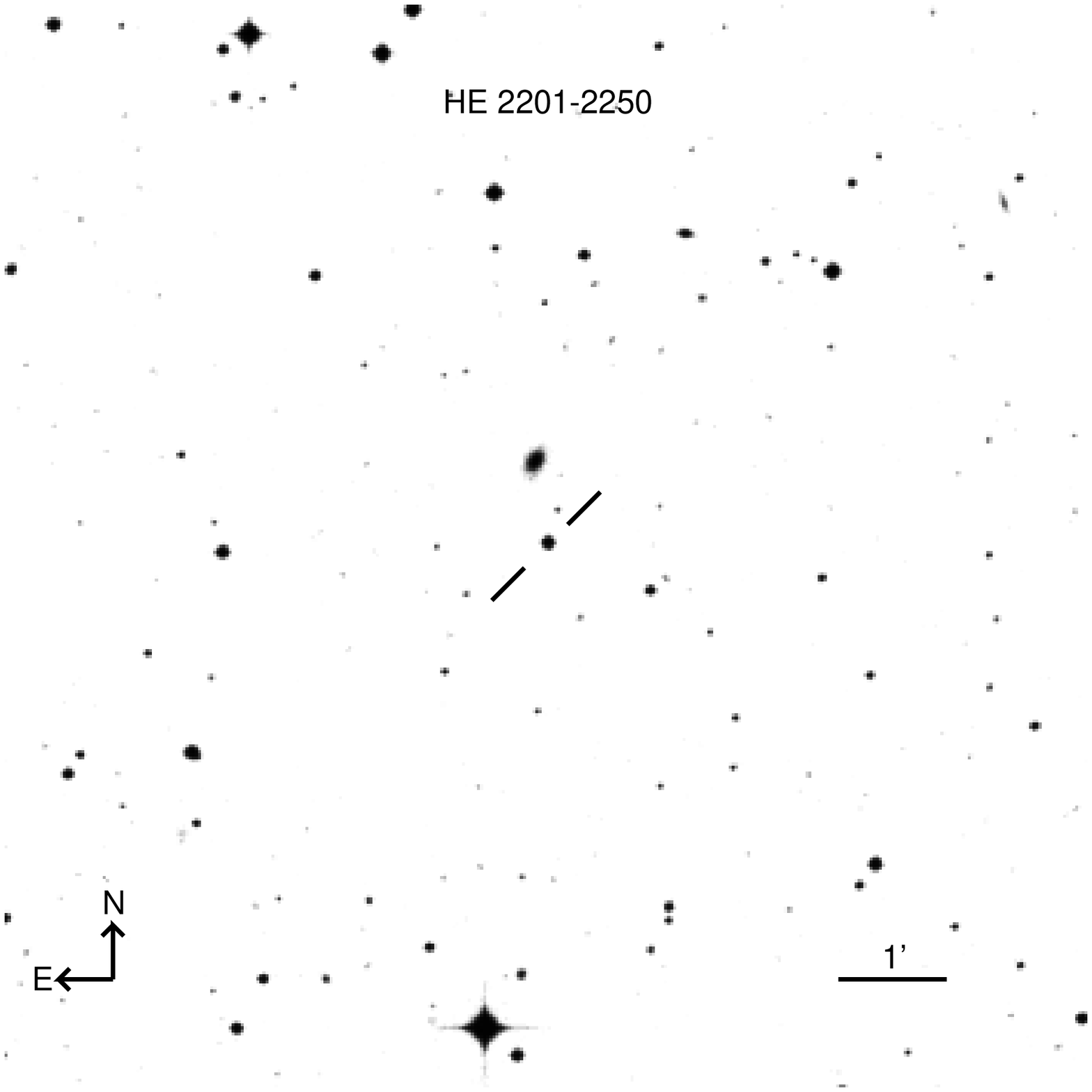}}}

\vspace{1.2mm}
\fbox{{\centering
\epsfxsize=8.5cm
\epsffile{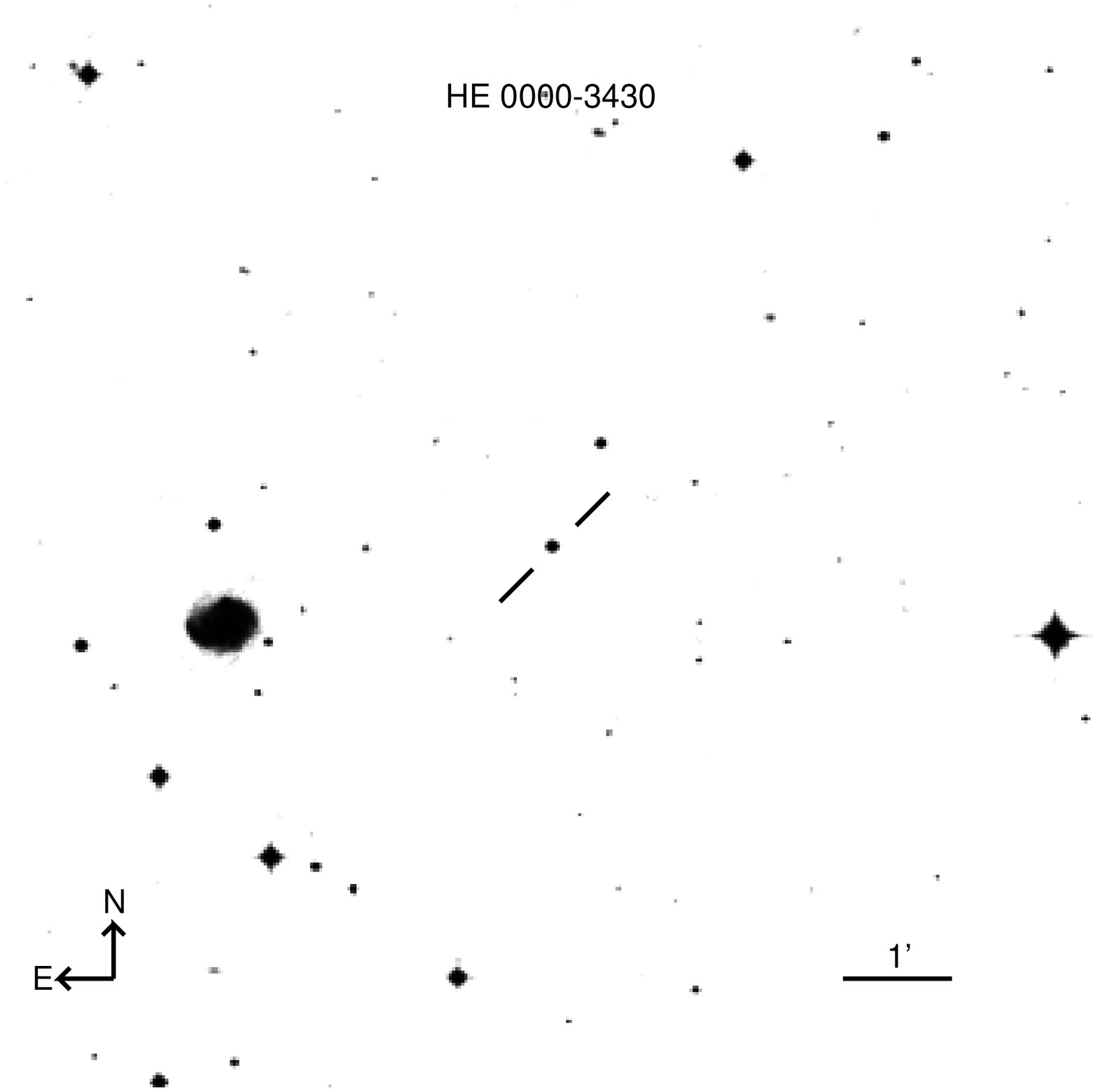}}}
\caption[]{Finding charts for HE\,2201-2250 and HE\,0000-3430
}
\label{ffc2}
\end{figure}  
\section{Observations}

\begin{table}[htb]
\caption[]{Coordinates and approximate magnitudes}
\begin{flushleft}
\begin{tabular}{llll}
\hline
Object & R.A. (1950) &  Decl. (1950)&  $V$ magnitude\\
\hline
HE\,1211-1707 &  12$^{\rm h}$11$^{\rm m}$ 33\hs 5 & -17$^{\circ}07^{\prime}58^{\prime \prime}$    & 16\fm 9$\pm$ 0\fm05 \\
HE\,0127-3110 &  01$^{\rm h}$27$^{\rm m}$ 37\hs 8 & -31$^{\circ}10^{\prime}34^{\prime \prime}$    & 16\fm 1$\pm$ 0\fm50 \\
HE\,2201-2250 &  22$^{\rm h}$01$^{\rm m}$ 57\hs 7 & -22$^{\circ}50^{\prime}51^{\prime \prime}$    & 16\fm 2$\pm$0\fm50  \\
HE\,0000-3430 &  00$^{\rm h}$00$^{\rm m}$ 06\hs 3 & -34$^{\circ}30^{\prime}07^{\prime \prime}$    & 15\fm 0$\pm$ 0\fm50 \\
\hline
\end{tabular}
\end{flushleft}  
\label{coo}
\end{table}

\begin{table}[htb]
\caption[]{Log of  observations}
\begin{flushleft}
\begin{tabular}{lllrr}
\hline
Object & Date in UT &  Instrument&  Resol.& Exp.\\
\hline

HE\,1211-1707 &  22.3.93, 7:28 &  1.5\,m+B\&C  &   6\,\AA & 60min\\
              &  12.1.94, 7:42 &  3.6\,m+EFOSC &  20\,\AA & 30min\\
              &  12.1.94, 8:02 &  3.6\,m+EFOSC &  20\,\AA & 15min\\
              &  12.1.94, 8:21 &  3.6\,m+EFOSC &  20\,\AA & 30min\\
              &  23.2.94, 1:54 &  2.2\,m+B\&C  &  12\,\AA & 60min\\
              &  01.3.95, 7:49 &  3.6\,m+EFOSC &  16\,\AA & 10min\\
              &  01.3.95, 8:11 &  3.6\,m+EFOSC &  16\,\AA & 10min\\
              &  02.3.95, 7:29 &  3.6\,m+EFOSC &  16\,\AA & 10min\\
              &  02.3.95, 7:51 &  3.6\,m+EFOSC &  16\,\AA & 10min\\
              &  02.3.95, 8:12 &  3.6\,m+EFOSC &  16\,\AA & 10min\\
              &  28.4.95        &  IUE SWP (LA) &  7\,\AA & 390min\\
              &  29.4.95        &  IUE SWP (LA) &  7\,\AA & 390min\\
HE\,0127-3110 & 24.09.94 & 1.5\,m+B\&C &  6\,\AA & 10min \\
              & 25.09.94 & 1.5\,m+B\&C &  6\,\AA & 30min \\
              & 23.11.94 & 1.5\,m+B\&C & 20\,\AA & 90min \\
HE\,2201-2250 & 23.11.94 & 1.5\,m+B\&C & 20\,\AA & 15min \\
HE\,0000-3430 & 20.11.94 & 1.5\,m+B\&C & 20\,\AA &  5min \\
\hline
\end{tabular}
\end{flushleft}  
\label{log}
\end{table}

\noindent
The four new magnetic degenerates have been discovered within the
Hamburg/ESO Survey for bright QSOs. The project is briefly
described by Reimers (1990) and Wisotzki  et al. (1995). The objects have
been selected as QSO candidates on the basis of their blue
continuum and the apparent absence of Balmer H or He lines.
Coordinates and approximate magnitudes are given in Table\,\ref{coo};
in Table\,\ref{log} a log of the observations is listed.

\begin{figure}[htbp]
\epsfxsize=8.8cm
\epsffile{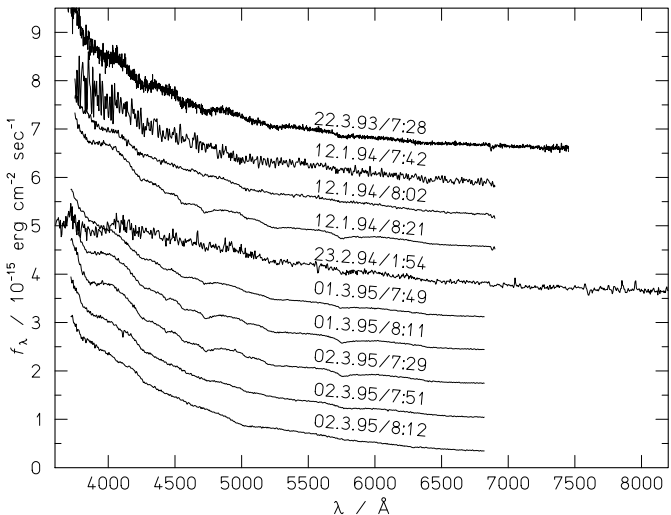}
\caption[]{Ten optical spectra of HE1211-1707 obtained at ESO,
a Silla. The zero point corresponds
to the lower spectrum; for clarity the other spectra are shifted
upwards. 
}
\label{fhe1211}
\end{figure}

\begin{figure}[htbp]
\epsfxsize=8.8cm
\epsffile{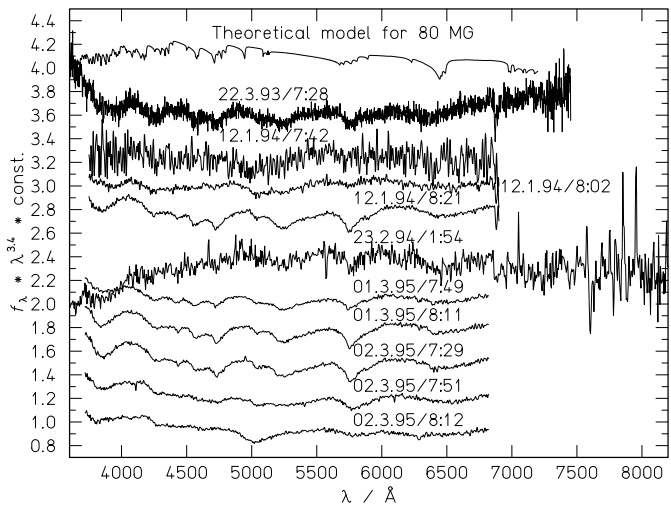}
\caption[]{Same spectra as in Fig.\ref{fhe1211}, but in order to
discern the weak features more clearly, the flux is ``normalized''
by multiplication of $f_{\lambda}$ by $\lambda^{3.4}*$const.
}
\label{fhe1211n}
\end{figure}

HE\,1211-1707:
From the newly discovered magnetic white dwarfs HE\,1211-1707 is
certainly the most peculiar object.
Optical spectroscopy with  the 1.5\,m Telescope at ESO  in
March 1993 reveiled a  spectrum with broad features, none of which could
be identified with any known spectral line. Subsequent observations in January
1994 were even more puzzling: within 20 minutes the features dramatically   
changed
their position and strength. This behaviour was also confirmed by
 EFOSC observations performed in March 1995
(see Fig.\ref{fhe1211} and Fig.\ref{fhe1211n}). 
Three of the spectra (12.1.94, 8:21UT; 01.3.95, 8:11UT;
02.3.95, 7:29UT) are extremely similar and are obviously taken during
the phase where the features are strongest. In the other spectra the
absorption bands are  shallower and at different positions. The lower
three spectra taken within one hour show a time sequence where the
the strong  depression at 5700\,\AA\ becomes weaker, whereas a shallower
absorption gradually becomes stronger at 5000\,\AA; this feature is also
visible in two other spectra taken on January 12, 1994.
A further spectrum that
covers in particular the red spectral range up to 8200 \AA\ has been
taken in February 1994 at the Calar Alto 2.2\,m telescope with the
aim to detect possibly present $\sigma$ components of H$\alpha$. The
spectrum is, however, much too noisy to identify additional
absorption bands. In the blue  the spectrum resembles those taken
during the ``strong feature'' phase.

\begin{figure}[htbp]
\epsfxsize=8.8cm
\epsffile{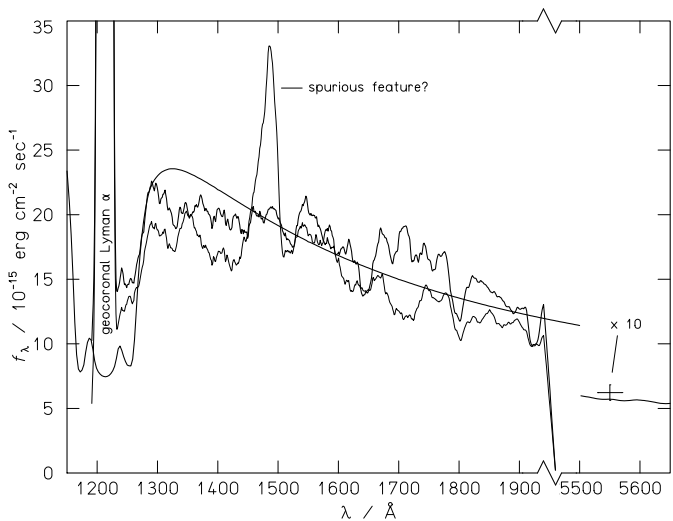}
\caption[]{
Two  very low S/N ($\approx$ 3) UV spectra of HE1211-1707,
taken with the IUE satellite
and smoothed with a boxcar of 10\,\AA. The strong and broad emission line
at 1485\,\AA\ is probably spurious since it only occurs in one exposure.
The IUE spectrum and the optical flux (the latter is calculated from
the $V$ magnitude) is compared with  a  80\,MG  dipole model with
$T_{\rm eff}=23000$\,K. For clarity the optical flux is multiplied by
a factor of ten
}
\label{fhe1211iue}
\end{figure}

Fig.~\ref{fhe1211iue} shows two UV spectra taken with
the SWP camera of the IUE satellite on April 28 and 29,  1995.
The signal-to-noise ratio is about 3 making the interpretation rather
problematic.
The strong and broad emission line at 1485\,\AA\ is probably an artifact since it is
only present  in one exposure of the exposures.

\begin{figure}[htbp]
\epsfxsize=8.8cm
\epsffile{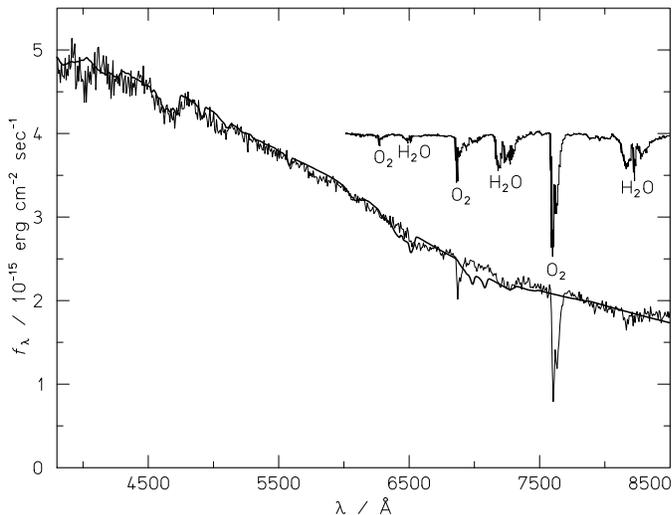}
\caption[]{Spectrum of HE\,0000-3430, taken with the ESO 1.5\,m telescope,
 compared to a synthetic spectrum for
a magnetic dipole with polar field strength of 86 MG, offset by -0.1
stellar radii toward the southern magnetic pole. The magnetic axis of
the model is inclined at $-45^{\circ}$ to the line of sight, the effective 
temperature is of the model is 7000\,K. The red part of the spectrum is 
strongly blended by  telluric lines. For comparison a spectrum from 
Stevenson (1994) is shown where the atmospheric bands are identified
}
\label{fhe0000}
\end{figure}  

\begin{figure}[htbp]
\epsfxsize=8.8cm
\epsffile{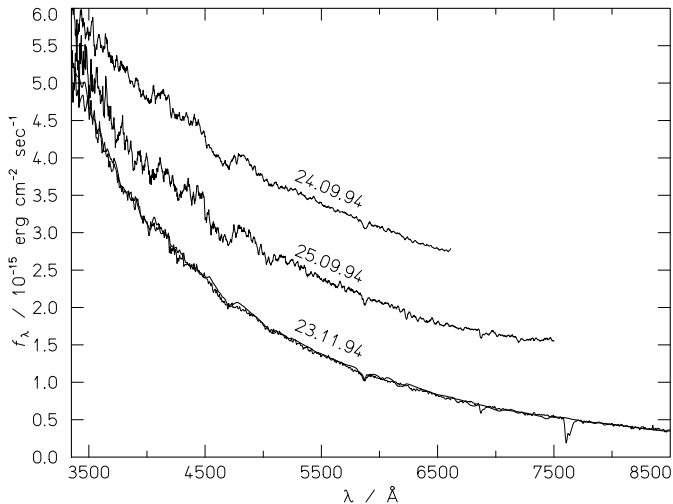}
\caption[]{Three spectra of HE\,0127-3110 taken with the ESO 1.5\,m
telescope.
The zero point corresponds to the high S/N  spectrum taken on November 23, 1994
; for clarity the other
spectra are shifted upwards. Note that the upper too spectra are
smoothed with a boxcar of 10\,\AA\ width and were taken under 
non-photometric conditions.
The lower spectrum is compared to an offset-dipole  model (-0.2 stellar
radii)  with a polar field strength of 176\,MG. The inclination of the
dipole axis is $-50^{\circ}$ and $\Teff=18000$\,K. Note that the strength
of the H$\beta$ absorption at 4650\,\AA\ is somewhat stronger in the
September 1994 spectra, probably due to rotational modulation 
%stronger
}
\label{fhe0127}
\end{figure}  

\begin{figure}[htbp]
\epsfxsize=8.8cm
\epsffile{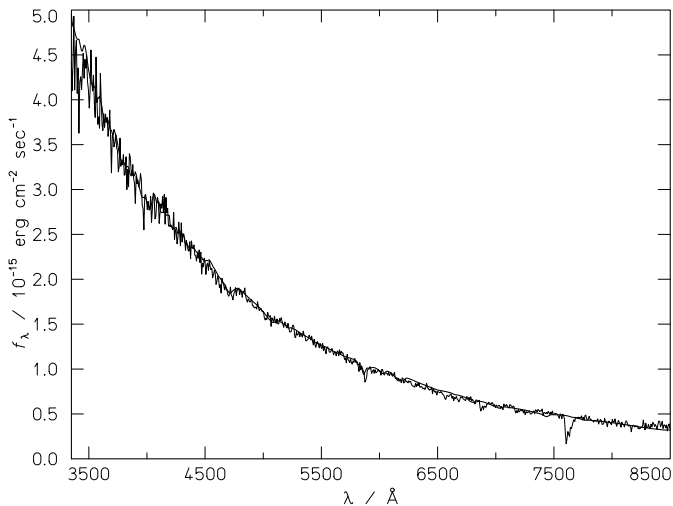}
\label{fhe2201}
\caption[]{The spectrum of HE\,2201-2250, taken with the ESO 1.5\,m telescope,
 is compared to the same model as its ``twin'' HE\,0127-3110 (November 23, 1994,
observation). Within the  noise level no significant differences can be found.
}
\end{figure}  

\begin{figure}[htbp]
\epsfxsize=8.8cm
\epsffile{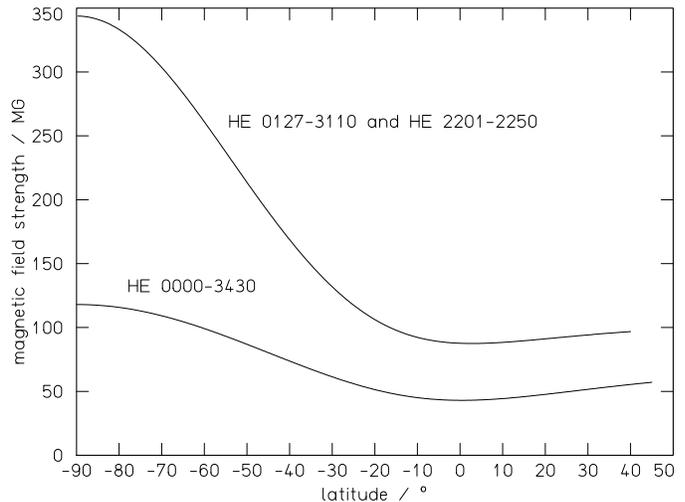}
\label{fhemagfield}
\caption[]{Distribution of the magnetic field strength vs. latitude for
the best-fit models used in Fig.~\ref{fhe0000}-\ref{fhe2201}
}
\end{figure}  

HE\,0127-3110: two spectroscopic observations have been made in September
1994 using the B \& C spectrograph at the ESO 1.5\,m telescope. Its
spectrogramm is characterized by broad bands around 4000, 4300,
4700, and 5100 \AA\ and a much sharper feature at $\approx 5850$ \AA\ 
(Fig.\ref{fhe0127}). A new spectrum with higher S/N has been taken on
Nov. 13, 1994 with the 1.5\,m telescope. The absorption features appear weaker
than in the Sept. 94 spectra.

HE\,2201-2250 has been observed in November 1994 also at the ESO
1.5\,m telescope. It is virtually a twin to HE\,0127-3310 both in its
spectral energy distribution and absorption lines.

HE\,0000-3430: the slope of the spectrum taken with the ESO 1.5\,m telescope 
is much smaller compared to the spectra of the other three objects. This 
indicates a much lower effective temperature.  The most prominent feature
is an absorption band at about 4650 \AA.

\section{Interpretation and Discussion}

A large grid of about 800 theoretical spectra for different effective 
temperatures and magnetic field configurations have been calculated
with a program developed in Kiel (see Jordan 1992 and Putney \&\ Jordan 1995
for a detailed description). For this analysis the magnetic  field
 configuration  
is described by dipoles shifted by 0., 0.1, 0.2 or 0.3 stellar radii
along the magnetic axis. The viewing angle $\alpha$ is defined such that
$+90^\circ$ and  $-90^\circ$ means that the observer is directly looking at
the  north and south magnetic pole, respectively, and 
$0^\circ$ corresponds to an  equator-on view. The third free parameter is
the magnetic field strength at the pole $B_{p}$ of the unshifted dipole.

Zero field model atmospheres 
 (see Koester et al. 1979) were used to determine the
 temperature and pressure structure  of the white dwarf's  outer layers;
convection has been suppressed since the magnetic field substantially
decreases the convective efficiency.
 The gravitational acceleration
at the surface was assumed to be $10^8$~cm\ts sec$^{-2}$.
The four equations of radiative transfer %(Beckers 1969)
were solved for the
 Stokes parameters $I$, $Q$, $V$, and $U$ on a large number of surface
elements (about 5000))
 with a procedure
similar to  the  semi-analytical algorithm described by Wickramasinghe \&\
Martin (1979), and subsequently   combined in order
to obtain the result from the
whole visible hemisphere of the star.
We considered line absorption by  all Balmer components up
to H$_{\epsilon}$
(Forster et al. 1984, R\"osner et al. 1984, and additional data from
the T\"ubingen group). The magnetooptical parameters  were treated  according 
to Jordan  et al. (1991). The  photoionization cross sections
were  determined by  using shifted zero field values according to
Lamb \& Sutherland (1974). Only minor changes are expected for the
flux spectra when the newly
calculated bound-free opacities by the Bochum group (Merani et al. 1995) are
used instead of the Lamb \&\ Sutherland approximation (Jordan \&\ Merani 1995
 and in prep.).

\subsection{HE\,1211-1707}

This object is by far the most interesting of the four objects due
to its  exceptionally fast variation of the spectral features.
In general  white dwarfs have turned out to be slow rotators (e.g. Koester 
and Herrero 1988) and there are several  magnetic white dwarfs like 
Grw$ +70^{\circ}8247$ whose  spectropolarimetry is constant on time scales
 up to at least 10 years (Schmidt \&\ Norsworthy 1991). Within about 20
minutes the clearly visible absorption features apparently disappear 
almost completely. 

The most probable explanation for this behaviour is that HE1211-1707
is a magnetic white dwarf. Although the faintness of this object
($V=16.9$) did
not allow a measurement of the polarization, the
spectral variations are similar to that of a  a few other magnetic objects,
e.g. in PG1031+234, the white dwarf with the highest known
field strength. For this star Schmidt et al. (1986)), and
Latter et. al. (1987) found a rotational period of
3.4 hours; they concluded that in  addition to a global dipole field
with a polar field strength of 500\,MG a region (spot?) must exist
with a field up to 1\,GG.

 Recently, Barstow,  Jordan et al.
(1995) discovered that the hottest (50kK) known magnetic
white dwarf RE\,J\,0317-853 has a rotational period of 725 seconds, and
a rather inhomogeneous magnetic field varying from about 170\,MG to almost 
660\,MG ($B_{p}$=340\,MG, offset offset by 0.2 stellar radii toward the
southern magnetic pole). The rotational period of  RE\,J\,0317-853 could
be measured with the help of high speed optical photometry; it turned out
that the amplitude is $\approx \pm0\fm1$. Photometric measurements of
HE\,1211-1707 has been performed by Darragh O'Donoghue. However, no 
significant optical variation has been found.
From the speed of the spectral variation we can only estimate the 
rotational period to lie between that of PG1015+014
(98.7 minutes; Wickramasinghe \&\ Cropper 1988)
and RE\,J\,0317-853. 

In order to understand the spectra of HE1211-1707 we
have calculated a large grid of high-temperature magnetic white
dwarf model atmospheres. However, we have not yet been able to identify the
visible features unambiguously. The absorption features between 4000 and
5000~\AA,
visible during the ``strong feature'' phase (see Fig.~\ref{fhe1211n})
have a strong similarity with theoretical profiles of H$\beta$ and
H$\gamma$ at about 80~MG. The same is true for the absorption at
5700~\AA, which at these field strengths would be caused by
blue shifted H$\alpha$ components.
There is, however, no possibility to explain
the absorption at $\lambda \approx 5300$~\AA\  at  such a low field,
a feature which is present in the same spectra. Therefore we are
not sure if there are indeed regions on the surface of the star,
with a field as low a 80~MG.

Although we believe that this is true 
it is also possible that similar features may occur at
field strengths larger than 1000\,MG, hitherto the limit of our model
calculations (currently the line data sets for the model atmospheres are
completed to account for  magnetic fields extending 1\,GG).
One  possibility is a global dipole-like field of about 80\,MG plus
regions on the star with a considerably larger magnetic field strengths.
This would mean that the field is at least as complicated as 
PG1031+234. Another alternative is 
that the star belongs to the group of
highly magnetic white dwarfs with unidentified absorption features
most probably containing helium in their atmospheres (e.g. GD229,
Greenstein et al. 1974). However, Engelhard \&\ Bues
(1995) have proposed that the
hitherto unexplained features in GD229 are not due to helium but
quasi-Landau resonances at extremely high fields ($> 10^9$\,G). 
It would be interesting to test this hypothesis for the  case of
 HE1211-1707 as well.

UV observations have turned out to be rather
helpful to determine the magnetic field strength, because this spectral
region shows one or two of the three Lyman $\alpha$ components.
The presence of these absorptions would
prove that hydrogen is the dominating absorber in HE1211-1707. 
The position of the red shifted  Lyman $\alpha$ $\sigma$ component
would directly tell us the dominating field strength, as has been shown
by Barstow et al. (1995) for RE\,J\,0317-853.

Unfortunately,
the star is much too faint for the  IUE spectrum (Fig.\,\ref{fhe1211iue},
S/N$\approx 3$) to answer this question unambiguously.
We believe, however, that the UV spectrum confirms our speculation that
a  part of the surface is dominated by a field strength of about 
80\,MG: the absorption trough close to the
geocoronal Lyman$\alpha$ emission is so broad, that it cannot be explained
by the Lyman$\alpha$ $\pi$ component only (see Fig.\,\ref{fhe1211iue}).
 Therefore it is probably
blended  by  the red shifted  $\sigma$ component  meaning that the 
magnetic field responsible for this width  must be lower than about
100\,MG. A UV spectrum taken with HST is necessary to test if this is
really the case and would also help to find additional absorption features
produced by the  Lyman$\alpha$  $\sigma$ component at other field strengths.

Since the slope of the optical spectra varies strongly from observation
to observation due to non-photometric weather conditions it is difficult
%sj: 17.2 -> 16.9
to determine an effective temperature. Using the $V$ magnitude of $16\fm 9$
and the  the flux level of the IUE spectra we could, however, constrain the
effective temperature to about 20000~K and 25000\,K
(see Fig.\,~ref{fhe1211iue})
so that HE\,1211-1707 is one of the hottest  magnetic white
dwarfs known.

\subsection{HE\,0127-3110}

The spectrum shown in Fig. (Fig.\ref{fhe0127}) is characterized by broad 
absorption
features around 4000, 4300, 4650 and 5080 \AA\ respectively and a
narrow feature at $\approx 5870$ \AA.

At a first glance one might tentatively identify the 5870 \AA\ line
with \ion{He}{I} 5875 \AA\ and the two longest wavelength features with Swan
bands of C$_2$ as in DO stars. However, there is no 4471 \AA\ \ion{He}{I} line,
no 5500 \AA\ band (one of the Swan bands), and the two bands
near 4000\,\AA\ and 4300\,AA\ remain unidentified.
A more convincing identification of the 5870\,\AA\ feature - a similar
line is seen in PG\,1031+234 - is that of a stationary 2s0$\rightarrow$3p0
 H$\alpha$ component at magnetic field strengths of about 230\,MG.

We have compared the November 1994 spectrum to theoretical  spectra
assuming a centered magnetic dipole. None of the models was able to reproduce
the observed spectrum in detail. All absorption features are shallower
than observed and also in some cases at the wrong positions. This indicates
that the variation of the magnetic field is larger than in the case of a
centered dipole, where the range of field strength covers a factor of two.
 The next attempt was to assume dipoles offset  by 0.2 stellar radii toward
the southern magnetic pole. The best solution was found for  a polar field 
strength of 176\,MG seen and a viewing angle of $\alpha=-50^{\circ}$
(see Fig.~\ref{fhe0127}). For such a model the field strength varies
between 85 and 345 MG (see Fig.\,\ref{fhemagfield}). Both the slope of the
continuum of the November 1994 spectrum  and the line strength are
compatible with an effective temperature of $\Teff=18000\pm 1000$\,K. 

The two spectra of HE\,0127-3110 taken in September have a rather low
signal-to-noise ratio. Therefore we were not able to perform a detailed
analysis with synthetic spectra. If smoothed with a boxcar of 10\,\AA\
one can see that the absorption features, especially at
4650\,\AA\ are somewhat stronger than in the November 1994 observation.
This is probably also due to rotation: probably the field looks 
somewhat more
homogeneous during the September 1994 observations (corresponding to a
larger viewing angle  $\alpha$) compared to the
spectrum taken in November, 1995. 
 In the future we hope
that time resolved spectra will enable us to determine the overall 
magnetic field configuration in some more detail.

\subsection{HE\,2201-2250}

It is very surprising that the spectrum of HE\,2201-2250 looks almost
identical to that of  HE\,0127-3110, especially since no simple dipole
configuration can explain the observations in detail. 
The similarity is so large that we could not find any significant differences
within the observational uncertainties: In Fig.\,\ref{fhe2201} we compared
the  spectrum of  HE\,2201-2250  to same model spectrum  that has been used
to explain the observation of HE\,0127-3110. Therefore we conclude that
the effective temperature and the approximate field distribution over the 
visible hemisphere must be about the same. Only time resolve spectroscopy
and polarimetry can prove if this also the case for the detailed field
structure. Note that the description of the field by a dipole offset along
the magnetic axis is only one way to account for the observations
with a minimum number of free parameters. This  offset provides
a reasonable description of the field variation (either smaller or larger
than in the case of a centered dipole). It can, however, not be excluded
that the stars have  offsets perpendicular to the magnetic axis or even more
complicated field geometries (e.g. global dipole-like structures plus a
magnetic spot). Putney \&\ Jordan (1995) have shown that one flux spectrum
of a magnetic white dwarf can be described by several slightly different
model geometries (offset dipoles or  dipole-quadrupole combinations)
while a measurement of the polarization strongly constraints the possible
field parameters. This means that despite the similarity of the
flux spectra of  HE\,2201-2250 and HE\,0127-3110, the star may still
have different field configurations in detail.

\subsection{HE\,0000-3430}

From the continuum slope we have estimated an effective temperature of
7000\,K. Although the error may be relatively large due to observational
uncertainties, the star is certainly the coolest white dwarf discovered by
the Hamburg/ESO survey. The reason is that  due to the Zeeman
splitting the depth of the  Balmer lines is strongly reduced compared to
the situation without a magnetic field. Therefore magnetic white dwarfs
have a  bluer continuum and are easier found by the selection criteria
of the Hamburg/ESO survey. The next hotter white dwarf found by the HE survey
is also magnetic (HE1045-0908, Reimers et al. 1994) while the coolest
non-magnetic DAs have temperatures above about 11000\,K.

The strongest feature in the spectrum of HE\,0000-3430 looks very similar
to the complex of H$\beta$ components in magnetic white dwarfs of 
moderate field strengths ($\lappr 100$\,MG). However, strong 
absorption by  H$\alpha$ is seen in the spectrum, and especially
the blue shifted $\sigma$ component of H$\alpha$ is rather shallow.
This, again, can only be explained if a spread of the field strength larger
than a factor of two is assumed: the observed spectrum is best reproduced
by a dipole offset by -0.1 stellar radii toward the southern magnetic pole
and a viewing angle $\alpha$ of $-45^{\circ}$. The corresponding range of
magnetic fields is 43\,MG to 118\,MG (see Fig.\,\ref{fhemagfield}).
 For such a field the H$\alpha$
$\pi$ components are shifted to about 6500\,\AA\ and while the 
$\sigma$ components are indeed rather shallow. Several components
of H$\beta$ are also well reproduced by the synthetic spectrum. Note, that
the spectrum is strongly blended by telluric lines at wavelengths above
6000\,\AA. For an easy identification of the atmospheric O$_2$ and
H$_2$O bands the average continuum-normalized spectrum of two metal-poor giant
stars, HD\,195636 and BD$-09^{\circ}$5831, taken from Stevenson (1994),
has been plotted in  Fig.\,\ref{fhe0000}  for comparison.

\section{Conclusions}

Beside quasars, the Hamburg/ESO Survey has turned out to be a rich source of
interesting blue stars. Together with HE1045-0908 the survey has enlarged
the number of magnetic white dwarfs with field strengths above 10\,MG
by 25\% (Schmidt \&\ Smith 1995).

With HE\,1211-1707  we have discovered one of the  hottest magnetic white
dwarfs showing a rather fast rotational modulation of the spectrum.
Most likely, the star has a rather complicated structure of the magnetic
field. The hemisphere that is seen during the rotational phase where the
absorption features are strongest is probably dominated by a magnetic field
of about 80\,MG, while much higher field strengths are present on the rest
of the star. Without better time resolved spectra (which are also necessary
to determine the rotational period)
and high S/N UV spectroscopy an unambiguous identification of the absorption
features remains impossible at
the moment. 

Recently, Muslimov et al. (1995) performed
theoretical calculations demonstrating that non-dipole components
can survive much longer on  the white dwarf cooling sequence
than previously believed. This is confirmed by our result that
all four stars
have a spread of magnetic field strengths larger than in the case of
a centered dipole-field. 

The spectra of HE\,0127-3110, HE\,2201-2250, and HE\,0000-3430 have been
successfully modelled with the help of synthetic spectra in the framework
of offset dipoles geometries. Putney \&\ Jordan (1995) have also
considered  dipole-quadrupole combinations as  a more physical  
model. However, for a detailed analysis which can distinguish between both
these geometries it would be necessary to obtain measurements of the
 polarization first. The flux spectrum alone can be reproduced
by  several slightly different offset-dipole models or
dipole-quadrupole combinations.

The fact that the spectra of many magnetic white dwarf show evidence
for non-centered magnetic dipoles strengthens the connection to Ap stars as
their precursors since these objects also exhibit large deviations from the
centered dipole structure.

Time resolved observations  of the rotating white dwarfs HE\,1211-1707 and
possibly HE\,0127-3110 will 
  allow to determine the field geometry over a large
portion of the stellar surface. This will provide a  rather important test
for  theoretical calculations of the magnetic field structure.

\acknowledgements
We are grateful to Darragh O'Donoghue for performing high-speed photometry for 
HE\,1211-1707. We also thank Hans-J\"urgen Hagen for taking a spectrum at the 
Calar Alto 2.2\,m telescope. The DFG partly supports the Hamburg/ESO survey
under Re\,353/33. Detlev Koester gratefully acknowledges a travel grant
(Ko 738/9-1) from the DFG.
For the finding charts we have made use of the Digitized Sky Survey 
which was  produced at the Space Telescope
Science Institute under US Government grant NAG W-2166.

%-------------------------------------------------------------------------------
\end{document}